\documentclass[conference]{IEEEtran}
\IEEEoverridecommandlockouts
\usepackage{cite}
\usepackage{amsmath,amssymb,amsfonts}
\usepackage{graphicx}
\usepackage{textcomp}
\usepackage{xcolor}
\usepackage{booktabs}
\usepackage{array}
\usepackage{url}
\def\BibTeX{{\rm B\kern-.05em{\sc i\kern-.025em b}\kern-.08em
    T\kern-.1667em\lower.7ex\hbox{E}\kern-.125emX}}

\begin{document}

\title{Appropriateness of Empathy in AI:\\ A Signal-Cost Perspective}

\author{
\IEEEauthorblockN{Chi-Ching Juan}
\IEEEauthorblockA{\textit{School of Information} \\
\textit{University of Toronto} \\
Toronto, Canada \\
sonia.juan@mail.utoronto.ca}
\and
\IEEEauthorblockN{Tao Wang}
\IEEEauthorblockA{\textit{School of Information} \\
\textit{University of Toronto} \\
Toronto, Canada \\
taotw.wang@utoronto.ca}
\and
\IEEEauthorblockN{Harold Lee}
\IEEEauthorblockA{\textit{Independent Researcher} \\
Toronto, Canada \\
harold.lee.kh@gmail.com}
\thanks{Preprint. Under review.}
}

\maketitle

\begin{abstract}
The appropriateness of empathy in AI has emerged as a critical concern,
as excessive empathy risks seeming manipulative while insufficient
empathy appears dismissive. While prior research has explored how to
quantify empathy in AI, few studies examine whether such empathy is
contextually appropriate. This paper introduces an economic perspective
by applying signaling theory to human--AI conversations. We propose
Signal Cost Proxies (emotional richness, perspective-taking, and
contextual tailoring) mapped to affective, cognitive, and associative
empathy. This multidimensional framework enables systematic evaluation
of empathy not just by presence, but by its appropriateness relative to
user demand.
\end{abstract}


\section{Introduction}
Ever since OpenAI acknowledged that their language models tended to
``over-please'' users in April 2025 \cite{ref1}, the appropriateness of
empathy has become an urgent topic. Exaggerated empathy risks coming
across as artificial or manipulative, while insufficient empathy can
feel cold or dismissive. Examining empathy's appropriateness is
therefore central to designing AI that communicates smoothly while
fostering trust and meaningful connection.

Existing studies of AI empathy primarily measure absolute levels using
psychological scales, linguistic metrics, or user ratings. Frameworks
such as the Multidimensional Evaluation Framework \cite{ref2},
LLaMA-EMRank \cite{ref3}, and ESHCC \cite{ref4} have advanced
operationalization across structural, behavioral, and perceptual
dimensions. The Illusion of Empathy \cite{ref5} further highlights the
gap between expressed and perceived empathy, underscoring the need to
assess appropriateness rather than quantity. Yet this line of research
remains limited by weak theoretical grounding, which fails to capture
empathy's affective, cognitive, and associative complexity. Meanwhile,
industry guidelines often advocate for ``lightweight empathy,'' however,
the notion remains vague and lacks systematic evaluation.

This work introduces an economic perspective to evaluate empathy in
LLMs. Drawing on signaling theory, we conceptualize empathy as a
cost-bearing signal: users express emotional demand through their
messages, and chatbots respond with varying signal costs to achieve
perceived appropriateness. This perspective reframes empathy not as a
fixed capacity but as a dynamic alignment between emotional expression
and contextual demand.

\section{Framework Development}
Cognitive science suggests that the human brain is predisposed to react
to signals, such as tone, phrasing, timing, and politeness, rather than
to the underlying identity of the message source \cite{ref6}. Applying
signaling theory to human--AI conversations, we interpret empathy not
only as a psychological construct but also as an economic signal, where
the credibility of empathetic responses depends on the relative cost
invested in communication. Accordingly, we operationalize empathy
appropriateness through three Signal Cost Proxies: emotional richness,
perspective-taking, and contextual anchoring.

We then map these proxies onto established psychological categories of
empathy: emotional richness aligns with affective empathy (sharing
emotions) \cite{ref7}, perspective-taking aligns with cognitive empathy
(understanding perspectives) \cite{ref8}, and contextual anchoring
aligns with associative empathy (relating to others' lived contexts)
\cite{ref9} (see Table~\ref{tab:proxies}).

\begin{table}[htbp]
\caption{Signal Cost Proxies for Empathy Appropriateness}
\label{tab:proxies}
\centering
\renewcommand{\arraystretch}{1.3}
\begin{tabular}{>{\raggedright\arraybackslash}p{1.5cm} >{\raggedright\arraybackslash}p{3.6cm} >{\raggedright\arraybackslash}p{1.6cm}}
\toprule
\textbf{Signal Cost Proxy} & \textbf{Evaluation Dimension} & \textbf{Empathy Type} \\
\midrule
Emotional Richness & Presence and strength of affective language, hedges, or soothing tone & Affective Empathy \\
Perspective Taking & Tracking through stance, epistemic markers, and perspective reciprocity & Cognitive Empathy \\
Contextual Tailoring & Use of details from the user's disclosure & Associative Empathy \\
\bottomrule
\end{tabular}
\end{table}

Emotional richness captures the intensity and nuance of affective
markers, with elaborated expressions signaling a greater investment in
the interaction. Perspective-taking reflects the effort to reframe or
validate the user's viewpoint, requiring cognitive engagement to
demonstrate genuine understanding. Contextual tailoring anchors empathy
in the specific circumstances of the conversation by incorporating
details from the user's disclosure, thereby grounding the response in
context (see Table~\ref{tab:examples} for more examples of high and low
signal cost).

\begin{table}[htbp]
\caption{Example of Signals with High and Low Cost}
\label{tab:examples}
\centering
\renewcommand{\arraystretch}{1.3}
\begin{tabular}{>{\raggedright\arraybackslash}p{1.9cm} >{\raggedright\arraybackslash}p{5.0cm}}
\toprule
\textbf{Signal Cost Proxy} & \textbf{Example of High and Low Signal Cost} \\
\midrule
Emotional Richness & High: ``I'm really sorry, that must feel overwhelming.'' \newline Low: ``Ok, I understand.'' \\
Perspective Taking & High: ``I can see why you may feel anxious with the exam tomorrow.'' \newline Low: ``Good luck on the exam.'' \\
Contextual Tailoring & High: mentions ``your mom's surgery'' or ``losing your job.'' \newline Low: generic ``that situation'' \\
\bottomrule
\end{tabular}
\end{table}

\section{Evaluation Methods Discussion}
To empirically test our proposed framework, we plan to examine the WASSA
2023 dataset \cite{ref10}, which includes paired human--chatbot
conversations annotated for empathy and related psychological variables.

The first stage of analysis focuses on quantifying empathy demand in the
human question. Here, emotional richness is measured by the presence and
intensity of affective vocabulary, the use of hedges or intensity
markers, and the degree of elaboration in the message \cite{ref11}.
Perspective-taking demand is captured when users disclose reasoning about
their feelings or invite others to consider their viewpoint. Finally,
contextual anchoring demand is reflected in the specificity of details
shared, for example referencing a family member's illness or a concrete
life event. Together, these three elements constitute the level of
communicative ``cost'' invested by the human, which we interpret as the
strength of their demand for empathy.

The second stage applies the same three dimensions to evaluate empathy
supply in the chatbot's reply. Emotional richness is observed in the
presence of nuanced affective vocabulary and elaborated expression, while
perspective-taking is demonstrated when the chatbot explicitly validates
or rephrases the user's viewpoint, for example, ``I can see why you'd
feel nervous about the exam.'' Here, perspective-taking is not assumed to
indicate genuine cognitive understanding. Instead, for chatbots, it is
operationalized as a linguistic proxy, defined as the ability to reframe
or validate the user's expressed viewpoint through language. In this
sense, ``showing understanding'' refers to observable linguistic
alignment rather than implying any internal cognition.

To avoid the chatbot's superficial mimicry, we further evaluate
perspective-taking along two dimensions: (i) semantic similarity between
the user's and chatbot's utterances (embedding-based alignment), and
(ii) perspective reciprocity, that is, whether the chatbot explicitly
acknowledges or responds to other's beliefs or emotions (e.g., A says
``I'm nervous about this,'' B replies ``I understand---it can be
stressful''). These criteria ensure that perspective-taking reflects
substantive alignment with the user's message rather than superficial
mimicry. Finally, contextual anchoring is measured by whether the chatbot
incorporates specific details from the human message into its reply. By
coding both sides of the exchange, we generate paired measures of demand
and supply.

The third stage computes a composite measure of appropriateness as the
match between empathy demand and supply. Importantly, this mismatch
judgment incorporates an adjustment factor, $\alpha$, to reflect
empirical findings that users systematically rate chatbots as less
empathetic than humans even when linguistic content is comparable. In
practice, this means that chatbot empathy supply is weighted down by a
fixed proportion (e.g., $X\%$) before mismatch is calculated.
Appropriateness is thus operationalized as
\begin{equation}
\mathit{Appropriateness} = 10 - \lvert \mathit{Demand} - \alpha \cdot \mathit{Supply} \rvert
\end{equation}
where $\alpha < 1$ adjusts chatbot empathy downward to align with
observed user perceptions.

This composite measure captures calibration rather than absolute supply,
emphasizing whether chatbot responses provide the right amount of empathy
relative to human expectations \cite{ref12}. To illustrate how this
measure works in practice, consider the following example. Suppose a user
writes ``I'm worried about my mom's surgery tomorrow.'' This signals high
contextual demand. A low-cost chatbot reply might be ``That situation
sounds stressful'' (low specificity), whereas a higher-cost reply would
be ``I can see why you'd feel anxious about your mom's surgery'' (high
specificity + semantic alignment). This example illustrates how the
framework captures appropriateness as the calibration between user demand
and chatbot supply. We expect that applying this method to WASSA 2023
conversations will reveal systematic under-supply of contextual anchoring
in chatbot responses.

\section{Conclusion}
This poster introduces a signal-cost framework for evaluating empathy
appropriateness in LLM-based chatbots. By framing empathy as an economic
signal, we propose three operational proxies: emotional richness,
perspective-taking, and contextual tailoring, which correspond to
affective, cognitive, and associative empathy, respectively. In future
work, we will apply this framework to annotated datasets to validate its
ability to capture nuanced trade-offs in empathetic communication.

\end{document}